# Convolutional Restricted Boltzmann Machine Based-Radiomics for Prediction of Pathological Complete Response to Neoadjuvant Chemotherapy in Breast Cancer

Li WANG, Lihui WANG, Qijian CHEN, Caixia SUN, Xinyu CHENG, Yuemin ZHU

***Abstract*— We proposed a novel convolutional restricted Boltzmann machine (CRBM)-based radiomic method for predicting pathologic complete response (pCR) to neoadjuvant chemotherapy treatment (NACT) in breast cancer. The method consists of extracting semantic features from CRBM network, and pCR prediction … It was evaluated on the dynamic contrast-enhanced magnetic resonance imaging (DCE-MRI) data of 57 patients and using the area under the receiver operating characteristic curve (AUC). Traditional radiomics features and the semantic features learned from CRBM network were extracted from the images acquired before and after the administration of NACT. After the feature selection, the support vector machine (SVM), logistic regression (LR) and random forest (RF) were trained to predict the pCR status. Compared to traditional radiomic methods, the proposed CRBM-based radiomic method yielded an AUC of 0.92 for the prediction with the images acquired before and after NACT, and an AUC of 0.87 for the pretreatment prediction, which was increased by about 38%. The results showed that the CRBM-based radiomic method provided a potential means for accurately predicting the pCR to NACT in breast cancer before the treatment, which is very useful for making more appropriate and personalized treatment regimens.**

***Index Terms*—Breast Cancer, DCE-MRI, Convolutional Restricted Boltzmann Machine, Pathological Complete Response, Neoadjuvant Chemotherapy**

## I. Introduction

BREAST cancer is the most common malignant tumor for women. More than 2 million new cases of invasive breast cancer were diagnosed and 626 thousand patients were killed by breast cancer in 2018 [1],[2]. The early diagnosis and precise treatment are very important for increasing the survival rate of breast cancer patients.

Traditional treatments for breast cancer in early stage include surgery, radiotherapy, and adjuvant systemic chemotherapy, etc. [3]. To decrease the rate of the cancer metastasis and increase the survival rate, in recent years, neoadjuvant chemotherapy treatment (NACT) has been used for patients with breast cancer [4]. It was demonstrated that a good response to NACT enables the size of the tumor to be reduced, which might increase the chance of breast-conserving surgery, allow eradicating micro-metastatic disease, and be useful to test the effectiveness of drugs used for the adjuvant chemotherapy [5]. Patients with pathological complete response (pCR) after NACT have a better prognosis and longer survival rate than those with non-pCR [6], [7]. Although most patients respond positively to NACT, showing a pathological complete response, there are still a majority of patients who remain resistant to NACT [8]. Therefore, predicting pCR and non-pCR status before NACT is very important to help doctors to make precise and personalized treatment regime for different patients, especially to help those non-responders to NACT to avoid ineffective therapies and missing the best treatment opportunity.

Conventionally, pCR status is examined from the surgical specimen of patients after the initial NACT completion in a histological manner, or evaluated with mammography and ultrasonography. With the development of imaging technology, dynamic contrast-enhanced magnetic resonance imaging (DCE-MRI) has become a promising imaging modality to assess the pCR after NACT due to its high sensitivity to angiogenic variations in the tumors [9]. The researches demonstrated that, after a short period of NACT, the changes in the kinetic parameter of DCE-MRI is highly related to pCR prediction [10]. Although these methods can provide an early response prediction following the initial NACT administration, it is unfortunately not possible to avoid ineffective treatment for the non-responders. Therefore, predicting accurately the pCR in an noninvasive manner before NACT is highly desired.

Recently, with the emerging of radiomics that attempts to characterize cancer properties by extracting high-throughput of quantitative features from multimodal images, more and more works about pCR prediction in breast cancer based on radiomics have been reported. For instance, Sun et al. extracted several DCE-MRI parameters from the images scanned after the first cycle of NACT to establish pCR prediction model using a logistic regression method; the prediction accuracy can

This paragraph of the first footnote will contain the date on which you submitted your paper for review. This work is funded partially by the National Nature Science Foundations of China (Grant No. 61661010), the Nature Science Foundation of Guizhou province (Qiankehe J No.20152044), the Program PHC-Cai Yuanpei 2018 (N° 41400TC) and the Guizhou Science and Technology Plan Project (Qiankehe [2018]5301). (Corresponding author: Lihui Wang.)

Lihui Wang is with Key Laboratory of Intelligent Medical Image Analysis and Precise Diagnosis of Guizhou Province, College of Computer Science and Technology, Guizhou University, Guiyang, 550025 (Email: lhwang2@gzu.edu.cn, wlh1984@gmail.com).

Li Wang, Qijian Chen, Caixia Sun, Xinyu Cheng are with Key Laboratory of Intelligent Medical Image Analysis and Precise Diagnosis of Guizhou Province, College of Computer Science and Technology, Guizhou University, Guiyang, 550025.

Yuemin Zhu is with CNRS UMR5220; Inserm U1206; INSA Lyon; Université de Lyon, France.

be up to 0.91 [11], which validates the potential of DCE-MRI for early prediction of pCR in breast cancer. Following this work, researchers tried to extract a large number of image textural features instead of kinetic features to predict pCR to NACT in breast cancer. Michoux et al. used gray-level co-occurrence matrix (GLCM) and gray-level run-length matrix (GLRLM) features to predict the pCR to NACT [12], the AUC of which can be up to 0.70. Giannini et al. extended the GLCM and GLRLM features into three-dimension (3D) to get a higher AUC of about 0.72 [13]. They succeeded in increasing the AUC to 0.84 by using several feature selection algorithms [14]. Considering that multimodal images could provide much more information, Liu et al. predicted the pCR with multimodal images for the patients who received NACT for 4 cycles, 6 cycles and 8 cycles, respectively [15].

These researches demonstrated that radiomics is a powerful means for predicting the pCR status to NACT. However, almost all the prediction methods mentioned above are based on the imaging data acquired after the administration of NACT. As far as we know, only a few works have investigated pCR prediction with the images acquired before NACT. Lin et al. predicted the PCR status with a higher AUC of 0.91 by combining the background parenchymal enhancement (BPE) and radiomic features extracted from the images acquired before NACT [16]. Plecha et al. further improved the prediction AUC by extracting radiomic features from both intratumoral and peritumoral regions in the peak enhanced images acquired before NACT [17]. Although they are able to get a higher prediction AUC, the requirements for image acquisition is a little strict. How to use fewer regularly acquired image data to get a higher prediction AUC is still a challenge.

Taking into account the advantages of unsupervised learning models in feature learning, in this work, we proposed to use convolutional restricted Boltzmann machine (CRBM) to extract semantic features instead of traditional radiomic features to increase pCR prediction accuracy with few images, especially for the pCR prediction before NACT. To evaluate the performance of the proposed method, the prediction accuracies obtained with traditional radiomic features and CRBM features are compared quantitatively.

## II. Materials and Methods

### A. Data Description

The dataset used in this study is named ISPY-1, which includes the DCE-MRI data of 222 patients [18]-[20]. It can be downloaded from the website of https://wiki.cancerimaging archive.net/display/Public/ISPY1. Considering that some slices do not have the labeled region of interest (ROI), 12176 slices with ROI were selected and taken as our dataset, which contains 4038 pCR samples and 8138 non-pCR samples. Breast DCE-MRI examination for each patient was performed respectively before administration of NACT (baseline), 1-3 days after NACT (early treatment), between anthracycline cyclophosphamide treatment and taxane therapy (inter-regimen), and after the final chemotherapy treatment and prior to surgery (pre-surgery). All the images were acquired with a 1.5 T scanner with a 3D fat-suppressed gradient echo sequence, in which, TR≤20 ms, TE = 4.5 ms, flip angle ≤45º, the field of view is about 16 to 18 cm, the minimum matrix is 256x192 with in-plane spatial resolution being less than 1 mm, and the slice thickness is less than 2.5 mm. The images were acquired once before contrast injection and repeated at least twice following injection.

Besides the imaging data, the clinical and pathological characteristics of the patients are given in TABLE I, including the patients' number, age, cancer subtypes, and imaging stages. Among these pathological characteristics, HER2 is an important factor for choosing the treatment strategy of breast cancer. Therefore, we compared the prediction accuracy of different methods for patients with HER2+ status. In addition, to compensate the insufficiency of current pCR prediction, we further divided the imaging data into four groups according to imaging stages, including baseline, early treatment, inter-regimens, and pre-surgery.

TABLE I
Clinical and pathological characteristics of different study populations.

| | pCR | Non-pCR |
|---|---|---|
| **Number of patients** | 17 | 36 |
| **Number of slices** | 4038 | 8138 |
| **Age (years)** | | |
| Mean | 44.16 | 47.81 |
| Std | 6.69 | 9.6 |
| **Imaging stages and subtypes** | | |
| **Baseline** | | |
| HR+,HER2- | 4 | 23 |
| TN/HER2+ | 13 | 13 |
| **Early treatment** | | |
| HR+,HER2- | 3 | 20 |
| TN/HER2+ | 13 | 11 |
| **Inter-regimens** | | |
| HR+,HER2- | 4 | 22 |
| TN/HER2+ | 11 | 12 |
| **Pre-surgery** | | |
| HR+, HER2- | 4 | 23 |
| TN/HER2+ | 13 | 13 |

NB: pCR denotes the pathological response to NAC, and Non-pCR the no response to NAC.

### B. Traditional Radiomic Pipeline for PCR Prediction

The traditional process of radiomics consists of four main steps. Firstly, the image ROI was delineated automatically or manually by radiologists. Secondly, a large number of human-defined image features were extracted, including first-order image features [21], [22] such as mean intensity, variance, skewness and kurtosis, the geometrical features (shape, volume, length, etc.), textual features (gray level co-occurrence matrix—GLCM and gray-level run-length matrix—GLRLM.) [23], and wavelet features [24]. Feature extraction programming was implemented in Matlab 2016a. A total of 335 quantitative features were automatically extracted from delineated ROIs to describe tumor phenotype characteristics. Thirdly, considering that the extracted features may be correlated and redundant, which will induce problems such as overfitting and poor generalization for the prediction model, feature selection was performed after the feature extraction. There are numerous feature selection algorithms; in the present work, we used partial least squared(PLS) method [25] to reduce feature dimension. Finally, based on the selected features, a classifier is employed for predicting. Currently, logistic regression (LR) [26], support vector machine (SVM) [27], random forest (RF) [28] are typical algorithms for classification. Therefore, these three algorithms were used for predicting pCR status. The overview of radiomic frameworks for pCR prediction is given in Fig. 1.

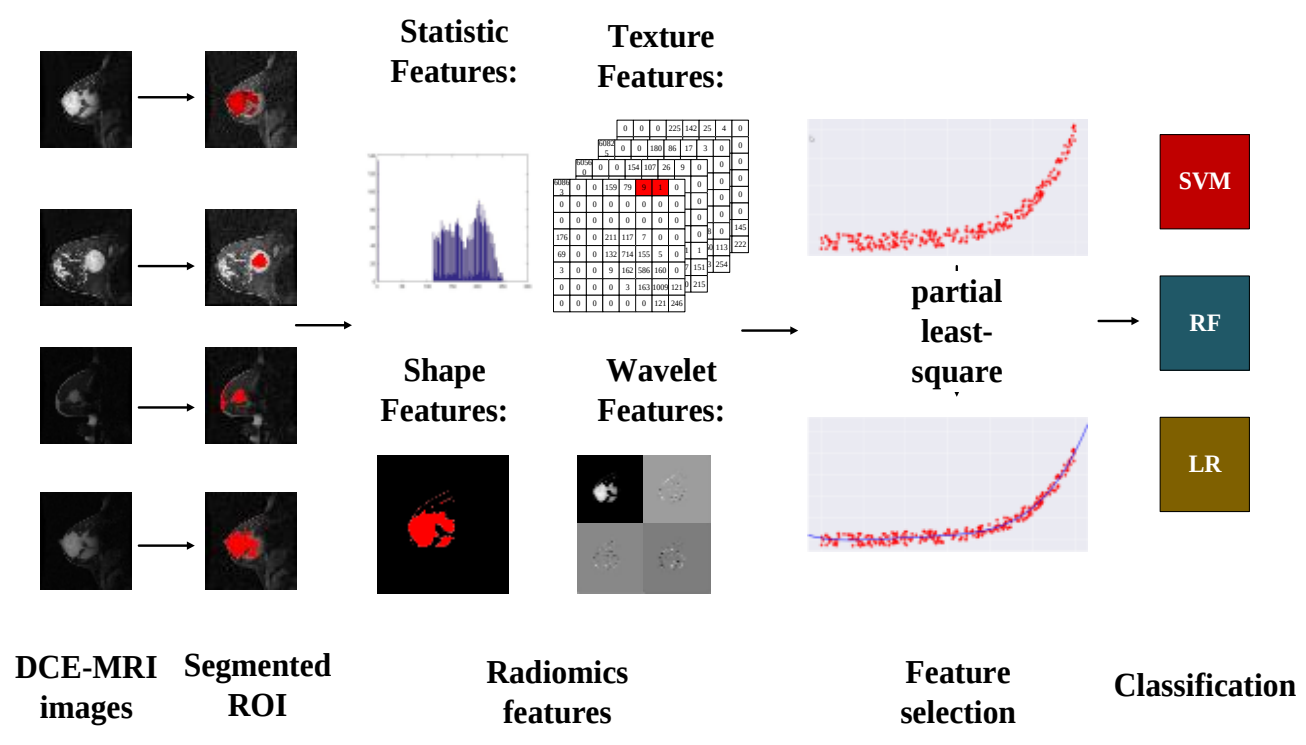

Fig. 1. Framework of traditional radiomics for pCR prediction.

## C. *Radiomics Based on Convolutional Restricted Boltzmann Machine for PCR Prediction*

As mentioned above, the predefined features in traditional radiomics may be insufficient to represent tumor phenotype characteristics and consequently influence the prediction accuracy. To deal with this issue, we propose to use CRBM to extract more semantic features. The framework of the CRBM features-based prediction is illustrated in Fig. 2.

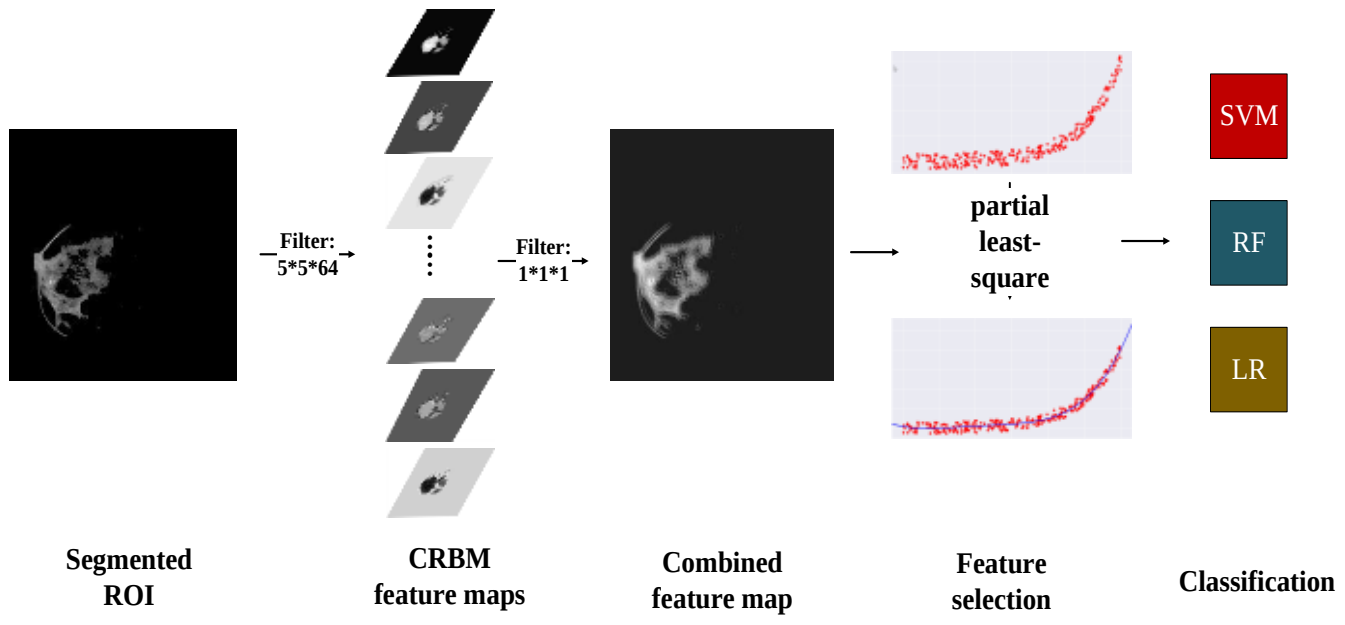

Fig. 2. Framework of CRBMs for pCR prediction.

With the segmented image ROI, some CRBM feature maps were obtained by the convolutions with convolutional kernels trained by CRBM method. Then, a convolution with kernel size of 1*1 were adopted to combine some features and reduce dimensions of CRBM feature maps. For the last two steps, we implement the same feature selection and classification method as traditional radiomics features based method.

The standard restricted Boltzmann machine (RBM) is an undirected probability graph model that contains a visible layer $v$ and a hidden layer $h$, where each unit in the visible layer is fully connected to the hidden units, but the units in the same layer are not connected, as illustrated in Fig. 3. RBM is devoted to learning the probability distribution of visible and hidden unites. In the application of image feature extraction, the units in the visible layer can be viewed as image pixels and those in the hidden layer as image features.

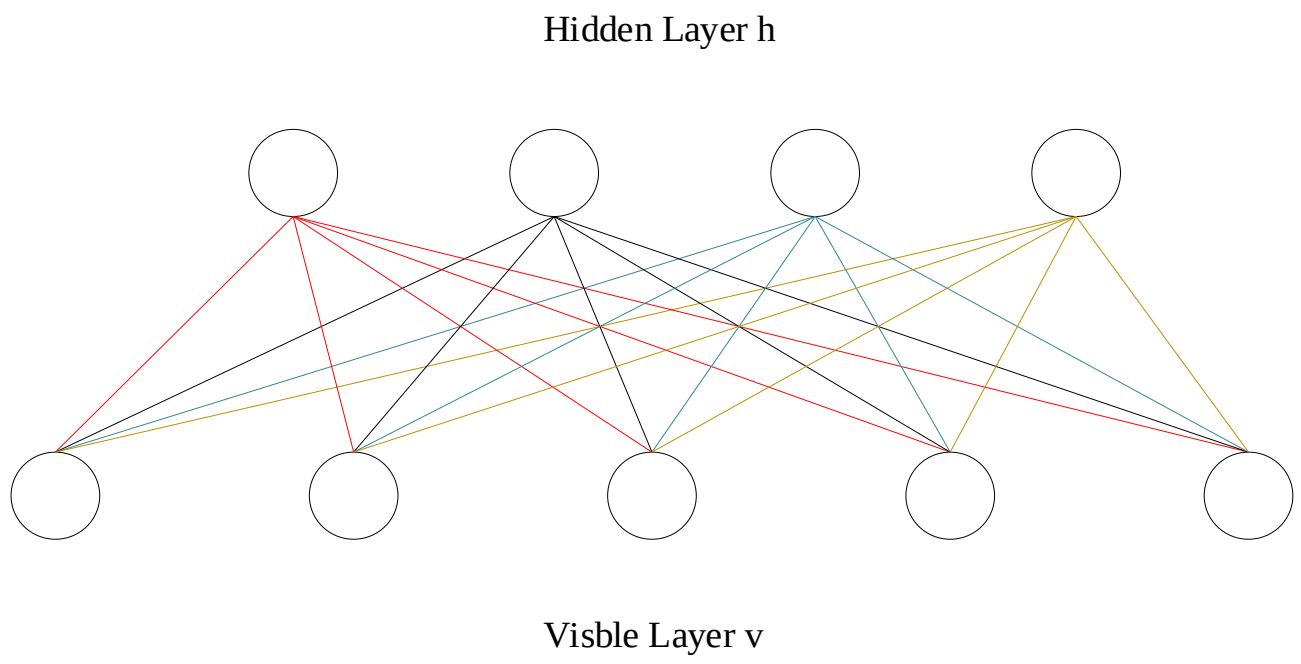

Fig. 3. Schematic diagram of standard RBM model.

RBM is an energy-based model, which means that the probability of variables in RBM is determined by the joint energy of hidden and visible units $E(\mathrm{v},\mathrm{h})$, namely,

$$P(v,h)=\frac{\exp(-E(v,h))}{Z}\ , \tag{1}$$

where $Z$ is a normalization constant defined by:

$$Z=\sum_{v}\sum_{h}\exp(-E(v,h)). \tag{2}$$

In the present study, the visible units and hidden units are both binaries. Thus, the energy $E(v,h)$ can be expressed as:

$$E(v,h)=\sum_{i}b_i v_i-\sum_{i,j}v_i W_{i,j}h_j-\sum_{j}c_j h_j\ , \tag{3}$$

where $v_i$ and $h_j$ represent respectively the $i^{th}$ and $\mathrm{j}^{th}$ units in visible and hidden layers, $b_j$ and $c_j$ are the bias items for units $v_i$ and $h_j$, and $W_{ij}$ is the connection weights between $v_i$ and $h_j$.

The probability inference from visible layer to hidden layer or vice versa can be formulated as:

$$\begin{aligned}P(h_j\mid \mathrm{v})&=sigm(\sum_{i}W_{i,j}v_i+c_j)\\ P(v_i\mid \mathrm{h})&=sigm(\sum_{j}W_{i,j}h_j+b_i)\end{aligned}\ , \tag{4}$$

where $sigm(x)=1/(1+\exp(-x))$ is the logistic sigmoid function.

From the RBM structure, we can observe that all the units in the visible layer are related to all the units in the hidden layer. Such dense connection is not practical for extracting features from big images. To overcome this problem, convolutional RBMs (CRBMs) were proposed, which benefit the weight sharing merits from convolution operations [29], the structure of which is depicted in Fig. 4.

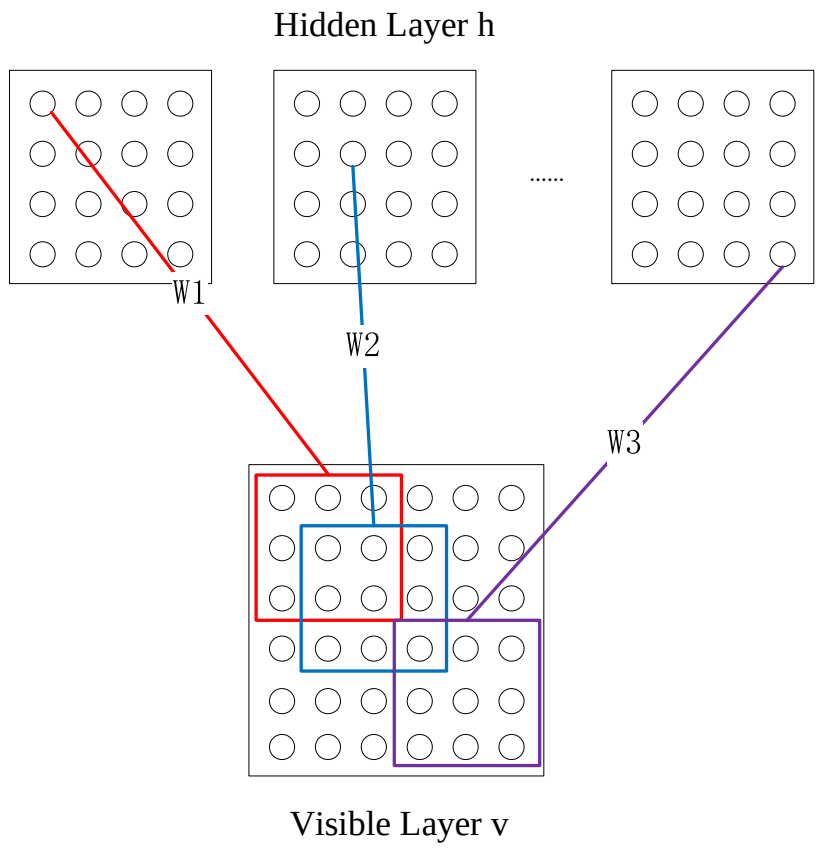

Fig. 4. Schematic diagram of convolutional RBM model.

We can see that the connections between hidden and visible layers are realized by several convolutional kernels $W_m$, which results in a hidden layer representing the image feature map $h_m$. If all the pixels in the input image share one bias $b$ and all the pixels in the $m^{th}$ hidden feature map share the same bias $c_m$, the probability inference can be rewritten as:

$$\begin{aligned} P(h_{ij}^m \mid \mathrm{v}) &= sigm((\tilde{W}_m * v)_{ij} + c_m) \\ P(v_{ij} \mid \mathrm{h}) &= sigm((\sum_m \tilde{W}_m * h_m)_{ij} + b) \end{aligned} , \quad (5)$$

where $\tilde{W}_m$ is the horizontally and vertically flipped filter of $W_m$, $h_{ij}^m$ denotes the pixel value located at $i^{th}$ row and $j^{th}$ column of the $m^{th}$ hidden feature map, and $v_{ij}$ represents the corresponding pixel value of the input image.

The purpose of RBM or CRBM is to find a set of parameters $\theta = \{W, b, c\}$ that minimizes the overall energy $E(v,h)$ with respect to the training data. Thus, the feature maps generated from the hidden layer can represent the input images perfectly. According to (1), minimizing the energy is equivalent to maximizing the log-likelihood function $L$:

$$\begin{aligned} L &= \ln P(v) \\ &= \ln(-\frac{1}{Z}\sum_h e^{-E(v,h)}) \\ &= \ln \sum_h e^{-E(v,h)} - \ln Z \\ &= \ln \sum_h e^{-E(v,h)} - \ln \sum_{h,v} e^{-E(v,h)} \end{aligned} . \quad (6)$$

The parameters are updated using the gradient descent method:

$$\theta := \theta + \eta \frac{\partial \ln P(v)}{\partial \theta} , \quad (7)$$

where $\eta$ indicates the learning rate. Eq. (6) shows that the calculation of $\sum_{h,v} e^{-E(v,h)}$ requires knowing the joint probability $P(v,h)$ which is unfortunately infeasible to infer. To cope with this problem, we used the contrastive divergence (CD) method of Hinton [30] , with which the updating of parameters can be written as:

$$\begin{aligned} W_{ij} &= W_{ij} + \eta( filter(v_i^0, P(h_j^0 \mid \mathrm{v}^0)) - filter(v_i^k, P(h_j^k \mid \mathrm{v}^k))) \\ b_i &= b_i + \eta(v_i^0 - v_i^k) \\ c_i &= c_i + \eta \sum_{j=1}^{n_h} (P(h_{i,j}^0 \mid \mathrm{v}^0) - P(h_{i,j}^k \mid \mathrm{v}^k)) / n_h \end{aligned} . \quad (8)$$

In (8), $v^k$ and $h^k$ are obtained by Gibbs sampling for k times. $n_h$ indicates the number of hidden nodes. Namely, given the input, the initial state of the hidden unit $h^0$ is sampled using the distribution $p(h^0 \mid v^0)$ as expressed in (5), where $v^0$ is the input image. Then, keeping the hidden unit unchanged, the status of visible units is updated as $v^1$ according to $p(v^1 \mid h^1)$. This alternating sampling process is repeated for k times to obtain $v^k$ and $h^k$. The detailed parameter updating algorithm is given in Algorithm 1 as follows.

**Algorithm 1** Stochastic parameter update of a binary CRBM

**Input:** $V^{(0)}$: Visible nodes; $W$: Filter wights;
$b$: Visible layer bias; $c$: Hidden layer bias;
$K$: Sampling numbers; $\eta$ : Learning rate;
**Initialization:**
$\Delta W = 0; \Delta b = 0; \Delta c = 0;$
$P(H|V)^{(0)} \leftarrow \sigma(filter(W, V^{(0)}) + b)$
$H^{(0)} \leftarrow Bernoulli(P(H|V)^{(0)})$
**Gibbs sampling iteration:**
**for** k=1,2,....,k **do**
$P(H|V)^{(k)} \leftarrow \sigma(filter(W, V^{(k-1)}) + b)$
$H^{(k)} \leftarrow Bernoulli(P(H|V)^{(k)})$
**end for**
**for** k=1,2,....,k **do**
$P(V|H)^{(k)} \leftarrow \sigma(filter(W^T, H^{(k-1)}) + c)$
$V^{(k)} \leftarrow Bernoulli(P(V|H)^{(k)})$
**end for**
**Updating CRBM parameters:**
**for** i=1,2,...,$n_h$;j=1,2,...,$n_v$ **do**
$\Delta W_{i,j} \leftarrow \Delta W_{i,j} + \eta(< P(H_i|V)^{(0)}, V_j^{(0)} >$
$- < P(H_i|V)^{(k)}, V_j^{(k)} >)$
$\Delta b_j \leftarrow \Delta b_j + \eta(V_j^{(0)} - V_j^{(K)})$
$\Delta c_i \leftarrow \Delta c_i + \eta(P(H_i|V)^{(0)} - P(H_i|V)^{(k)})$
**end for**
$W = W + \Delta W; b = b + \Delta b; c = c + \Delta c$ =0

## D. *Prediction Evaluation*

The performance of pCR prediction with traditional and CRBM-based radiomics was evaluated in terms of receiver operating characteristic (ROC) curve, area under curve (AUC), sensitivity, specificity, and accuracy. ROC is created by plotting true positive rate (TPR) against false positive rate (FPR) at various thresholds in a classifier. AUC is derived from ROC, which is the area under the ROC curve and specifies the classification accuracy. The bigger the AUC, the more accurate the classification. Sensitivity represents the correct classification rate of positive samples while specificity represents the correct classification rate of negative samples. Due to the fact that AUC is not sensitive to sample characteristics such as the unbalance of sample classes, it is often used to evaluate the performance of the classifier for the unbalanced dataset. The ROC and AUC are usually applied in dichotomous problem.

## III. Experimental Results

### A. Experimental Setup

In order to compare fairly the performance of traditional radiomics and CRBM-based radiomics for the prediction of pCR to NACT of breast cancer, all the experiments were implemented and evaluated with the same training and validation datasets. That is, the evaluation was performed using 4-fold cross-validation, in which 3044 samples were taken as the validation dataset and 9132 samples as the training dataset. Considering the fact that the image features extracted from CRBM are highly dependent on the input image size, in the comparison experiments, the input of CRBM was divided into two cases: one is the whole slice with a size of 256×256, and the other is the image patch with a size of 32×32.

The structure of CRBM used in the present work included one hidden layer, which was obtained by the convolutions between the input image and 64 convolutional kernels of size 5×5, and each convolution kernel had its own bias. During the training, the learning rate of CRBM was set as 1e-4. Under this setting, the dimension of feature maps derived from CRBM was 252×252×64 when the input was the whole slice and turned to 28×28×64 when the input was the image patch. To reduce the dimension and combine some features, a convolution with kernel size of 1×1 was performed, which resulted in a feature map of size 252×252 or 28×28. The feature map was flattened and taken as the input of classifiers. As to traditional radiomics, we extracted 335 features. Whatever the features extracted from CRBM or traditional radiomics, they were highly correlated. Therefore, a feature selection process was performed to reduce the number of features to 20.

### B. Quantitative Comparison of Overall PCR Prediction Accuracy

As mentioned above, the features that were inputted into the classifiers were divided into three groups: features defined in traditional radiomics, features extracted from image-based CRBM, and features from patch-based CRBM. To evaluate fairly the influence of image feature extraction methods on prediction accuracy, three classifiers including RF, SVM and LR were used. The ROCs of three classifiers for different image features are given in Fig. 5. In the ROC curves, the middle black curve represents the dividing line with AUC of 0.5, the blue curve indicates the ROC obtained with features from traditional radiomics, the green one the ROC from patch-based CRBM, and the orange the ROC from image-based CRBM. It can be seen that the prediction with traditional radiomic features has the worst performance for all the classifiers. The prediction accuracy using the features extracted from CRBM is much better than traditional radiomic features, especially for SVM and LR classifiers. Moreover, when these two classifiers are used, the prediction with the features extracted from patch-based CRBM is better than that with the features extracted from image-based CRBM. But for the RF classifier, the performance of patch-based CRBM is a little worse than that of image-based CRBM.

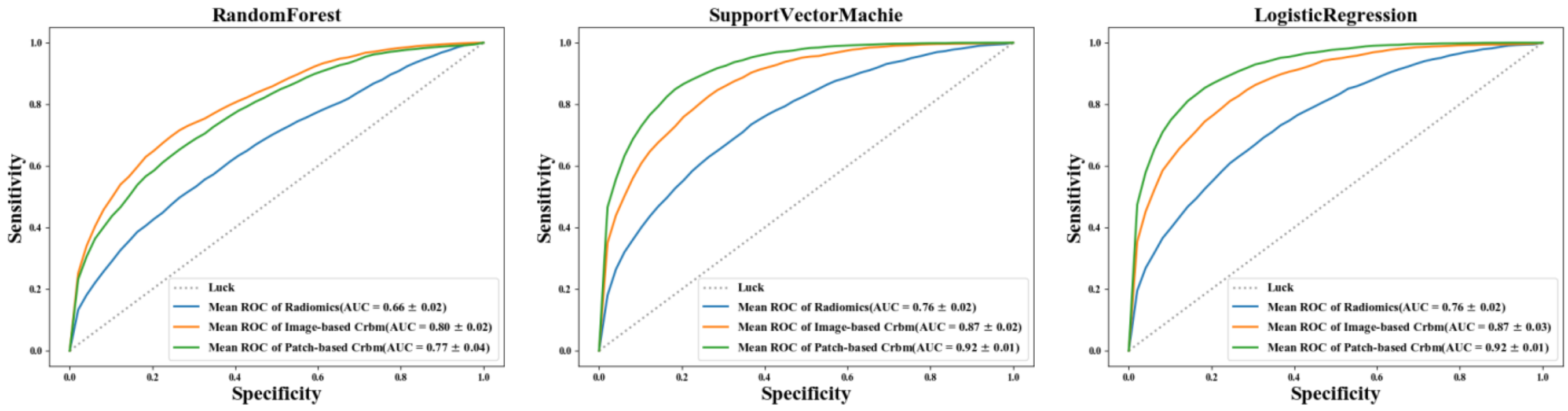

(a)RF (b) SVM (c) LR

Fig. 5. Comparison of ROCs for different classifiers with different image features.

To further quantitatively compare the pCR prediction performance with different features, the accuracy, sensitivity, specificity, and AUC were calculated and given in TABLE II. We observe that for most metrics, CRBM features led to better performance than traditional radiomic features. The prediction accuracy was increased by about 13% with SVM and LR, and 7% with RF.

TABLE II

QUANTITATIVE COMPARISON OF DIFFERENT CLASSIFIERS FOR PCR PREDICTION WITH DIFFERENT FEATURES IN TERMS OF VARIOUS METRICS

| | Accuracy | Sensitivity | Specificity | AUC |
|---|---|---|---|---|
| **LR** | | | | |
| Radiomics | 0.68 | 0.67 | 0.69 | 0.75 |
| Image-based CRBM | 0.76 | 0.82 | 0.73 | 0.87 |
| Patch-based CRBM | 0.83 | 0.83 | 0.83 | 0.92 |
| **SVM** | | | | |
| Radiomics | 0.67 | 0.72 | 0.64 | 0.75 |
| Image-based CRBM | 0.74 | 0.85 | 0.68 | 0.87 |
| Patch-based CRBM | 0.81 | 0.85 | 0.78 | 0.91 |
| **RF** | | | | |
| Radiomics | 0.65 | 0.51 | 0.73 | 0.66 |
| Image-based CRBM | 0.74 | 0.68 | 0.77 | 0.80 |
| Patch-based CRBM | 0.72 | 0.58 | 0.79 | 0.77 |

To better get insights into this effect, the feature discrepancy between positive samples (pCR) and negative samples (non-pCR) is given in Fig. 6, where the height of bars represents the feature discrepancy, the red and black points indicate respectively the feature values of positive and negative samples. It is seen that the discrepancy in features extracted form CRBM is much more obvious than that extracted from traditional radiomics. This explains well why the prediction based on the features extracted from CRBM is better than the traditional radiomics.

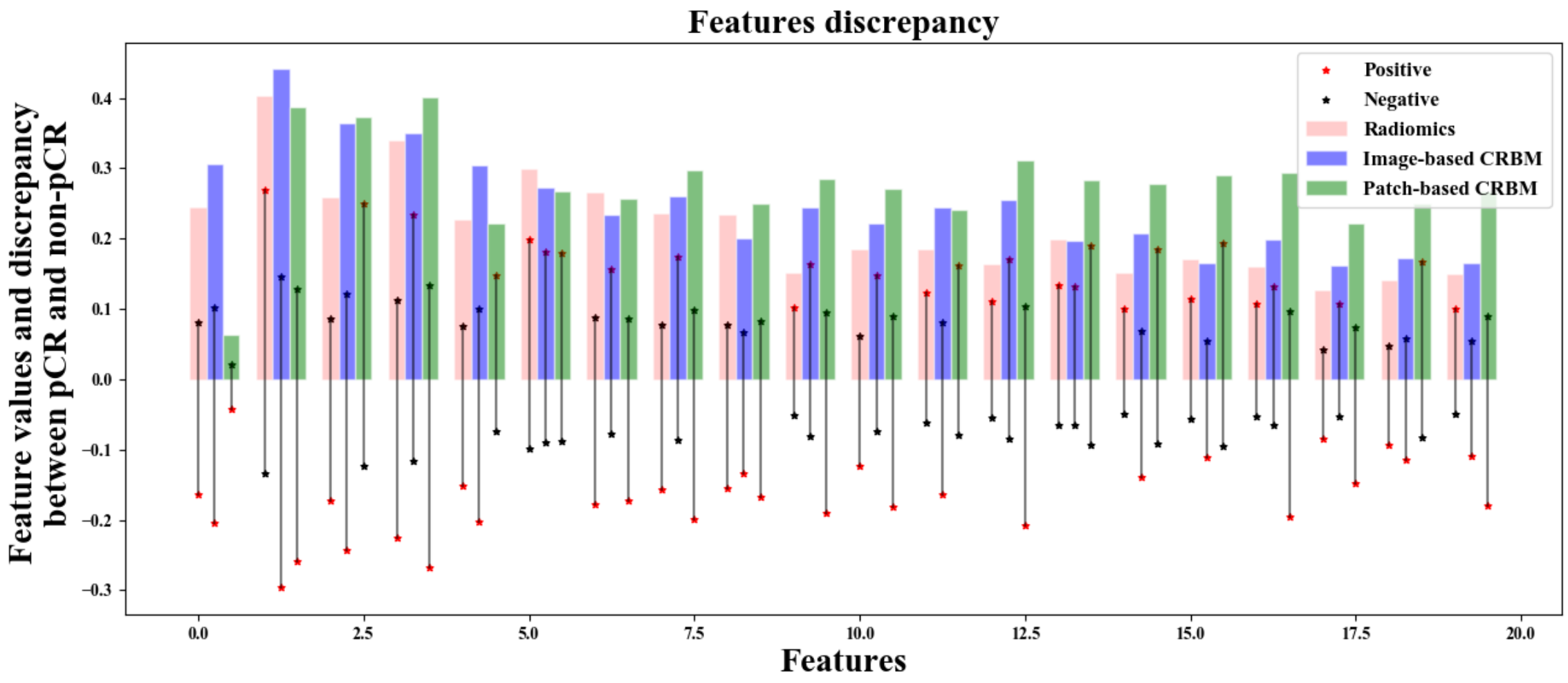

Fig. 6. Comparison of the feature discrepancies of different image features. The height of bars represents the feature discrepancy. The red and black points indicate respectively the feature values of positive and negative samples.

### *C. Quantitative Comparison of PCR Predictions Using the Images Acquired at Different Times*

The accuracy of pCR prediction is highly related to the breast cancer subtypes and imaging stages. Clinical experiments demonstrated that patients with different subtypes would respond to NACT in different ways. Generally, the patients with triple negative (TN) or human epidermal growth factor 2 positive (HER2+) breast cancer would get pCR more easily [31]. As to the imaging stages, the pCR prediction based on images acquired in baseline stage (before NACT) is much harder than using images acquired after NACT. That is why the pCR prediction before NACT was rarely reported. To further validate the performance of the proposed method, we divided the imaging data into five groups according to image acquisition times, and the corresponding predictions based on these images were respectively implemented. Considering the influence of data balance on prediction performance, only the images of the patients with TN and HER2+ breast cancer were used in this group of experiments. The LR classifier was chosen to compare the performance of features extracted from traditional radiomics and patch-based CRBM radiomics. The quantitative results were given in TABLE III.

TABLE III

QUANTITATIVE COMPARISON OF THE PREDICTIONS WITH THE IMAGES ACQUIRED AT DIFFERENT TIMES.

| | Accuracy | Sensitivity | Specificity | AUC |
|---|---|---|---|---|
| **All, HER2+/TN** | | | | |
| Radiomics | 0.66 | 0.68 | 0.64 | 0.73 |
| CRBM | 0.80 | 0.81 | 0.78 | 0.88 |
| **Baseline, HER2+/TN** | | | | |
| Radiomics | 0.58 | 0.58 | 0.59 | 0.63 |
| CRBM | 0.78 | 0.77 | 0.79 | 0.87 |
| **Early treatment, HER2+/TN** | | | | |
| Radiomics | 0.64 | 0.66 | 0.63 | 0.70 |
| CRBM | 0.71 | 0.65 | 0.79 | 0.78 |
| **Inter-Regimen, HER2+/TN** | | | | |
| Radiomics | 0.68 | 0.72 | 0.64 | 0.72 |
| CRBM | 0.72 | 0.71 | 0.74 | 0.81 |
| **Pre-surgery, HER2+/TN** | | | | |
| Radiomics | 0.60 | 0.63 | 0.56 | 0.66 |
| CRBM | 0.74 | 0.73 | 0.75 | 0.82 |

From TABLE III, it is observed that the accuracy of pCR prediction based on CRBM features was much higher than that based on traditional radiomic features for all the imaging stages. Especially at the stage of baseline, the AUC was increased by about 38%, which further validates that the semantic features extracted from CRBM are more appropriate for the pCR prediction before the initialization of NACT.

## IV. DISCUSSION

This work proposed a radiomic method to improve the prediction accuracy of pCR to NACT in breast cancer. The method is based on using CRBM to extract more semantic features. The outperformance of CRBM-based radiomics demonstrated that the semantic features extracted from the deep learning model are beneficial for the prediction of pCR to NACT in breast cancer, especially for the pretreatment prediction. The prediction accuracy of the proposed method was significantly improved compared to the models based on traditional radiomic features. Although some existing models could yield an AUC up to 0.8 [14] or even 0.9 [16], they are highly dependent on population size and imaging modalities. In contrast, when trained only with DCE-MRI images of about 40 patients, the proposed prediction model produced a better pCR prediction accuracy.

The experimental results demonstrated the importance of image features for the prediction using radiomics. Compared to conventionally defined features in the literature, such as shape, texture and geometrical features, the features learned from the CRBM network provide more representative information for the classification. In addition, thanks to the convolution operations, the features extracted from CRBM allow us to account for the correlation between different samples. This explains the better performance of the proposed method. Since the features extracted from CRBM were related to input image size, we analyzed the influence of the latter on the prediction accuracy. The experimental results showed that smaller patches convolved with small convolutional kernels yield better image features for the prediction.

Despite the high performance of CRBM-based features in predicting the pCR to NACT, there are still several limitations in the present work. Firstly, previous studies reported that the pCR prediction was related to receptor types, such as the status of HER2, HR and ER [32]. However, due to the unbalance of the dataset used (as indicated in Table 1) for the patients with HR+ and HER2-, namely, the number of pCR and non-pCR of this group are 4 and 23 respectively; the pCR prediction for such group was not analyzed. Secondly, it is well known that CRBM is sensitive to prior distribution; so considering the influence of different prior distributions would be interesting in the future. Thirdly, the validation of the proposed model was performed with the same cohort, which would degrade the statistical power of our model. In light of the merits of multiple imaging modalities, it would be useful to refine the proposed method to get more accurate and general pCR prediction models with more image datasets acquired from different centers.

## V. CONCLUSIONS

We proposed a novel CRBM-based radiomic method for predicting pCR to NACT in breast cancer before treatment. The method uses the semantic features extracted from CRBM instead of the human-defined features as done in traditional radiomics to predict pCR status. The experimental results demonstrated that the proposed radiomics achieved better performance than traditional radiomics, especially for the pretreatment prediction, the AUC of which is increased by about 38%. That suggests its use as a potential powerful tool for clinical pCR predictions to NACT in breast cancer.